\documentclass[twocolumn,showpacs,amsmath,amssymb]{revtex4}
\usepackage{graphicx}
\usepackage{dcolumn}
\usepackage{bm}
\usepackage{longtable}

\begin{document}

\title{
Nuclear Collective Tunneling Transitions \\
between Hartree States Deformed in Quadrupole Symmetry}

\author{M.~Maruyama}%
\affiliation{%
Department of Physics, Tohoku University, Sendai 980-8578, Japan 
}%
\author{T.~Kohmura}
\affiliation{ Department of Business Administration, 
Josai University, Sakado 350-0295, Japan 
}%
\author{Y.~Hashimoto}
\affiliation{%
Graduate School of Pure and Applied Sciences, 
University of Tsukuba, Tsukuba 305-8571, Japan 
}%

%
%

\date{\today}

\begin{abstract}
In a nucleus which has two Hartree states deformed in quadrupole symmetry, i.e., prolate state and oblate state, the nuclear residual interaction derived in the present theory beyond the Hartree approximation acts as the restoring force for the spherical symmetry of the nuclear system to be recovered so that the deformed nucleus makes collective tunneling transitions between prolate states and oblate states. 
We derive the Hamiltonian that is effective not only for the Hartree states but also for the collective tunneling transitions between the Hartree states. 
Solving the secular equation for the effective Hamiltonian on the basis states expressed in terms of SU(5) algebra, we analyze the nuclear collective tunneling transitions between a prolate state and an oblate state in the real-time description. 
The collective tunneling transitions are discussed to proceed through tri-axially asymmetric deformed states. \\
\end{abstract}

\pacs{21.60.-n}
\maketitle

\section{INTRODUCTION}

A nucleus is a self-contained system of strongly interacting nucleons, where the concept of Hartree is very useful to understand the mechanism of the system to be self-consistently bound: The Hartree approximation for a nucleus determines the nuclear shape and single-particle states. 
The nucleus of ${^{28}}$Si, for example, has two, i.e., prolate and oblate, deformed Hartree states. 
A Hartree state of an open-shell nucleus is characterized by the deformed Hartree potential, which is composed of the meson mean fields proper to the Hartree state \cite{Kohmura}\cite{Kohmura1}. 

The present theory beyond the Hartree approximation takes into account the fluctuation of the meson mean field proper to other Hartree states to take place in one Hartree state. 
The meson mean-field fluctuation proper to other Hartree states yields a residual interaction, which gives rise to the collective tunneling transitions from the Hartree state to another Hartree state in order to restore the spherical symmetry. 
Taking advantage of this picture of nuclear collective tunneling, in the previous papers \cite{Kohmura}\cite{Kohmura1}, we derived a Hamiltonian effective not only for deformed Hartree states but also for the collective tunneling transitions between the deformed Hartree states. 
The physical quantities in the Hamiltonian are determined by the Hartree calculations of the deformed states.

In this paper, we solve the secular equation for the nuclear effective Hamiltonian on the basis states expressed in terms of SU(5) algebra and analyze the collective tunneling transitions between the two deformed Hartree states in the real-time description in the present theory beyond the Hartree approximation. 
We discuss that the nuclear residual interaction, exciting $d$-state nucleons to $s$ state and vice versa, acts as the restoring force for the spherical symmetry of the nuclear system to be recovered so that the deformed nucleus makes collective tunneling transitions between prolate states and oblate states. 
The nucleus ${}^{28}$Si in the course of collective tunneling transitions between a prolate state and an oblate state proceeds through tri-axially asymmetric deformed states \cite{Bohr}\cite{Ring}. 

\section{HAMILTONIAN EFFECTIVE FOR HARTREE STATES 
AND FOR COLLECTIVE TUNNELING }

In the previous papers \cite{Kohmura}\cite{Kohmura1}, 
we derived a Hamiltonian effective not only 
for two deformed Hartree states but also for the collective tunneling 
transitions between the two deformed Hartree states. 
The physical quantities in the effective Hamiltonian 
are determined by the Hartree calculations of the deformed states.

Here we briefly review the formulation for the collective tunneling 
transitions between Hartree states in the description based 
on the non-relativistic meson mean-field calculations for the Hartree states. 
In the present description, for simplicity's sake, 
we adopt one sort of meson field in place of $\sigma$- 
and $\omega$-mesons and endorse the nuclear saturation by the truncation of the meson space. 
The non-relativistic nuclear field Hamiltonian is written as 
\begin{widetext}
\begin{eqnarray}
H_{\rm nucl}&=&\int[\psi^\dagger(\vec{r})(-{\nabla^2\over2M})\psi(\vec{r})+{1\over2}\{\pi^2(\vec{r})+\phi(\vec{r})(-\nabla^2+m^2)\phi(\vec{r})\} \nonumber \\ 
&&-g\phi(\vec{r})\psi^\dagger(\vec{r})\psi(\vec{r})]d^3r, \label{f}
\end{eqnarray}
\end{widetext}
where $\psi$ is nucleon field, $\phi$ and $\pi$ are meson fields, and $M$, $m$ and $g$ are  nucleon mass, meson mass and meson-nucleon coupling constant, respectively. 

We first solve the Hartree problems for a nuclear system 
which has two Hartree states $\Psi_i$ and $\Psi_f$. 
The two Hartree states are specified by the meson mean fields 
$\varphi^i$ and $\varphi^f$, respectively. 
In terms of the meson mean fields $\varphi^i$ 
and $\varphi^f$ in the two Hartree states determined 
in the Hartree calculations, 
we formulate the collective tunneling transitions 
between the two Hartree states. 
The meson mean field $\varphi$, which varies from $\varphi^i$ 
to $\varphi^f$ in the course of the collective tunneling, 
acts as a residual interaction to steer the nuclear collective 
tunneling transitions from $\Psi_i$ to $\Psi_f$ 
in the present theory beyond the Hartree approximation. 

We derive a Hamiltonian effective for the Hartree states 
and also for the collective tunneling transitions 
between the Hartree states \cite{Kohmura}\cite{Kohmura1}. 
In terms of the meson mean fields $\varphi^i$ 
and $\varphi^f$ in the two Hartree states $\Psi_i$ 
and $\Psi_f$, we define the tunneling steering field to be 
\begin{eqnarray}
\varphi_{\rm s}(\vec{r})={1\over2}\{\varphi^f(\vec{r})-\varphi^i(\vec{r})\}, \label{0} 
\end{eqnarray}
and the principal mean field to be 
\begin{eqnarray}
\varphi_{\rm p}(\vec{r})={1\over2}\{\varphi^i(\vec{r})+\varphi^f(\vec{r})\}. 
\end{eqnarray} 

Hartree states deformed in quadrupole symmetry, 
either prolate or oblate, are one of the features of the nuclei 
that have some valence nucleons in the $s$ and $d$ major shell. 
We formulate the collective tunneling transitions 
of $s$-$d$ shell nuclei between a prolate Hartree state 
and an oblate Hartree state. 
A typical example of the nuclei that exhibit both an oblate 
and a prolate Hartree state with an almost degenerate energy 
is the nucleus of ${}^{28}$Si \cite{Walet,Kohmura2,Lederer}, 
which has 6 protons and 6 neutrons in the $s$ and $d$ major shell. 
While a configuration of nucleons in the non-spherical $d$ states 
deforms the Hartree potential, the nuclear residual interaction that 
is of spherical symmetry gives rise to collective tunneling transitions 
between the prolate Hartree state and the oblate Hartree state.

Suppose that the nucleus ${}^{28}$Si is polarized along $z$-axis in the two, i.e., prolate and oblate, deformed Hartree states. 
The two Hartree states are symmetric to each other with respect to the deformation parameter \cite{Ohta}, i.e., the principal mean field in Eq. (3) is spherically symmetric, 
\begin{eqnarray} 
\varphi_{\rm p}(\vec{r})=\varphi_{\rm p}(r)Y_{00}(\theta,\varphi), 
\end{eqnarray} 
and the tunneling steering field in Eq. (2) is of quadrupole symmetry, 
\begin{eqnarray}
\varphi_{\rm s}(\vec{r})=\varphi_{\rm s}(r)Y_{20}(\theta,\varphi). 
\end{eqnarray}

We quantize the meson field $\phi(\vec{r})$ in terms of only the three essential fields, i.e., the principal mean field $\varphi_{\rm p}(\vec{r})$, the tunneling steering field of quadrupole symmetry, 
\begin{eqnarray}
\bar{\varphi}_{20}(\vec{r})=N_2\varphi_{\rm s}(r)Y_{20}(\theta,\varphi), 
\end{eqnarray}
and the other components of the quadrupole field, 
\begin{eqnarray}
\bar{\varphi}_{2m}(\vec{r})=N_2\varphi_{\rm s}(r)Y_{2m}(\theta,\varphi), 
\end{eqnarray} 
with the normalization factor $N_2$: 
\begin{eqnarray}
\phi(\vec{r})&=&\varphi_{\rm p}(\vec{r})+\hat{\phi}(\vec{r}), \\ 
\pi(\vec{r})&=&0, 
\end{eqnarray}
where the quadrupole field is quantized as 
\begin{eqnarray}
\hat{\phi}(\vec{r})=\sum_m{1\over\sqrt{2\omega_2}}\{a_{2m}\bar{\varphi}_{2m}(\vec{r})+a^\dagger_{2m}\bar{\varphi}^*_{2m}(\vec{r})\} 
\end{eqnarray}
and  
\begin{eqnarray}
\omega^2_2=\langle\bar{\varphi}_{2m}|-\nabla^2+m^2|\bar{\varphi}_{2m}\rangle. 
\end{eqnarray} 
The spherical principal mean field $\varphi_{\rm p}(\vec{r})$ 
is assumed to contribute only to the nuclear single-particle energies, 
but not to the residual interaction energies. 
The set of the quadrupole fields $\bar{\varphi}_{2m}$ 
make the meson field to be rotationally invariant. 
The components for $m\neq0$ of the quadrupole field 
are necessary to take into account any orientation of the deformed 
Hartree states in the initial and final states of the tunneling process 
and to give rise to the collective tunneling transitions 
between the Hartree states. 

The meson field $\phi$ in Eq. (8) is related to the nucleon field $\psi$ by the field equation, 
\begin{eqnarray}
(-\nabla^2+m^2)\phi(\vec{r})=g\psi^\dagger(\vec{r})\psi(\vec{r}). 
\end{eqnarray}
Projecting the above equation onto the quadrupole field, we express the quantum quadrupole meson field $\hat{\phi}$ in terms of the nucleon field: 
\begin{eqnarray}
\hat{\phi}(\vec{r})&=&{1\over\omega^2_2}\sum_m\bar{\varphi}_{2m}(\vec{r})\int\bar{\varphi}^*_{2m}(\vec{r'})(-\nabla'^2+m^2)\hat{\phi}(\vec{r'})d^3r' \nonumber \\
&=&{g\over\omega^2_2}\sum_m\bar{\varphi}_{2m}(\vec{r})\int\bar{\varphi}^*_{2m}(\vec{r'})\hat{\psi}^\dagger(\vec{r'})\hat{\psi}(\vec{r'})d^3r', 
\end{eqnarray}
where the nucleon field $\hat{\psi}(\vec{r})$ is quantized in terms of the creation and annihilation operators $c_{si}$ for $s$ nucleons and $c_{\mu i}$ for $d_\mu$ nucleons in the basis states determined in the single-particle potential for the spherical principal mean field $\varphi_{\rm p}(\vec{r})$ in Eq. (4): 
\begin{eqnarray}
\hat{\psi}(\vec{r})=\sum_i(c_{si}\psi_s(\vec{r})+\sum_\mu c_{\mu i}\psi_\mu(\vec{r})). 
\end{eqnarray}
The suffix $\mu$ stands for the $z$-component of the orbital angular 
momentum $l$=2 for $d$-state nucleons and the suffix $i$ 
for spin and isospin of nucleons. 

The expressions (8) and (12)-(13) for the meson field 
are substituted into the nuclear field Hamiltonian $H_{\rm nucl}$ 
in Eq. (\ref{f}). 
This procedure for the nuclear field Hamiltonian determines 
the Hamiltonian effective for the Hartree states 
and also for the collective tunneling: 
\begin{eqnarray}
H_{\rm nucl}=H_{\rm shell}+H_{\rm int}. \label{1} 
\end{eqnarray} 
The quadrupole meson fields $\bar{\varphi}_{2m}(\vec{r})$ 
yield the nuclear quadrupole-quadrupole interaction, 
responsible for the two deformed Hartree states 
and also for the collective tunneling transitions 
between the two Hartree states, 
\begin{widetext}
\begin{eqnarray}
H_{\rm int}=-{g^2\over2\omega^2_2}\int\int\sum_m\hat{\psi}^\dagger(\vec{r})\bar{\varphi}_{2m}(\vec{r})\hat{\psi}(\vec{r})\hat{\psi}^\dagger(\vec{r}')\bar{\varphi}^*_{2m}(\vec{r}')\hat{\psi}(\vec{r}')d^3rd^3r'. 
\end{eqnarray}
\end{widetext}
The shell-model Hamiltonian $H_{\rm shell}$ for nucleons 
in the $s$ and $d$ shells is for the spherical shell model, 
\begin{eqnarray}
H_{\rm shell}=\sum_{i}(\varepsilon_sc^\dagger_{si}c_{si}+\sum_\mu\varepsilon_dc^\dagger_{\mu i}c_{\mu i}). 
\end{eqnarray}
When the valence nucleon configurations 
are truncated in the $s$ and $d$ major shell, 
the quadrupole-quadrupole interaction Hamiltonian $H_{\rm int}$ 
in Eq. (16), responsible for the deformed Hartree states 
and also for the collective tunneling, is divided into the two
interactions, i.e., the interaction $H$ between a pair of $d$-state 
nucleons and the interaction $H_{sd}$ to excite $d$-state nucleons 
to $s$ state and vice versa, 
\begin{eqnarray}
H_{\rm int}=H+H_{sd}, 
\end{eqnarray}
where 
\begin{eqnarray}
H&=&-{g^2\over2\omega^2_2}\sum_{mm'ii'}u_{mm'}c^\dagger_{m+m'i}c^\dagger_{-m-m'i'}c_{-mi'}c_{mi}, \\
H_{sd}&=&-{g^2\over2\omega^2_2}\sum_{mii'}(v_mc^\dagger_{mi}c^\dagger_{-mi'}c_{si'}c_{si}+{\rm H. c.}). 
\end{eqnarray} 

The quadrupole-quadrupole interaction $H$ between a pair of $d$ nucleons plus the shell-model Hamiltonian $H_{\rm shell}$ form the deformed Hartree potential for the prolate Hartree state and also that for the oblate Hartree state. 
On the other hand, the $s$-$d$ interaction $H_{sd}$, which is spherically symmetric, yields the residual force for the prolate Hartree state to tunnel to the oblate Hartree state and for the return process in the present theory beyond the Hartree approximation. 
In the numerical calculations below, the single-particle energies $\varepsilon_s$ and $\varepsilon_d$ in the spherical shell-model Hamiltonian $H_{\rm shell}$ are assumed to be degenerate $(\varepsilon_s=\varepsilon_d)$ and the spin-orbit splitting for the $d$ states is ignored. 
In Section IV, we solve the secular equation for the nuclear Hamiltonian $H_{\rm nucl}$ in Eq. (15) and express the collective tunneling transitions between the two Hartree states in the real-time description. 
Taking advantage of SU(5) symmetry in the quadrupole-quadrupole interaction Hamiltonian $H$, we solve the problems of the collective tunneling. 
In the present theory, the $s$-$d$ interaction $H_{sd}$ plays role of the pairing interaction \cite{Bertsch, Bertsch2} in the collective tunneling between the deformed Hartree states. \\

\section{SU(5) EXPRESSION FOR INTERACTION HAMILTONIAN $H$ BETWEEN $d$-STATE NUCLEONS }

\subsection{ SU(5) expression for Hamiltonian $H$} 

The nucleus of ${}^{28}$Si, which has 12 nucleons 
in the $s$ and $d$ major shell, has a prolate and an oblate Hartree state. 
Suppose that the nucleus ${}^{28}$Si is polarized along $z$-axis in either a prolate or an oblate Hartree state. 
When the spin-orbit interactions are neglected, 
the single-particle states $(d_1\ d_{-1}\ d_0)^4$ 
are occupied in the prolate state and $(d_2\ d_{-2}\ s)^4$ 
are occupied in the oblate state, where $s$ and $d_\mu$ stand for 
the $s$ and $d$ single-particle states, respectively. 
The index 4 represents the degrees of freedom for spin and isospin for $s$ and $d$ nucleons. 

The transformation generators in the space of the five single-particle states $d_{\pm2},\ d_{\pm1}$ and $d_0$ constitute an SU(5) algebra. 
Actually the Hamiltonian $H$ in Eq. (19) is expressed in terms of the quadrupole operators $Q_m$ of the SU(5) generators, as is shown below. 
In this Subsection, we study the properties of the nuclear system governed by the Hamiltonian $H$ in Eq. (19) in terms of SU(5). 
The 24 generators for the SU(5) algebra are expressed by the creation and annihilation operators $c_{\mu i}$ for the $d$-state nucleons as the multi-pole operators: 
\begin{eqnarray}
L_m&=&\sqrt{10}\sum_{\mu i}(2\mu2m-\mu|1m)(-1)^{\mu-m}c^\dagger_{\mu i}c_{\mu-m i}, \\
Q_m&=&\sqrt{10}\sum_{\mu i}(2\mu2m-\mu|2m)(-1)^{\mu-m}c^\dagger_{\mu i}c_{\mu-m i}, \\
S_m&=&\sqrt{10}\sum_{\mu i}(2\mu2m-\mu|3m)(-1)^{\mu-m}c^\dagger_{\mu i}c_{\mu-m i}, \\
R_m&=&\sqrt{10}\sum_{\mu i}(2\mu2m-\mu|4m)(-1)^{\mu-m}c^\dagger_{\mu i}c_{\mu-m i}.
\end{eqnarray}
The $z$-component of the nuclear orbital angular momentum $L$, 
\begin{eqnarray}
L_0=\sum_i(2c^\dagger_{2i}c_{2i}+c^\dagger_{1i}c_{1i}-c^\dagger_{-1i}c_{-1i}-2c^\dagger_{-2i}c_{-2i}),
\end{eqnarray}
and the nuclear quadrupole moment, 
\begin{eqnarray}
-Q_0=-\sqrt{5\over7}\sum_i&\bigl(&2c^\dagger_{2i}c_{2i}-c^\dagger_{1i}c_{1i}-2c^\dagger_{0i}c_{0i} \nonumber \\
                          &-&c^\dagger_{-1i}c_{-1i}+2c^\dagger_{-2i}c_{-2i}\bigr), 
\end{eqnarray}
are additive collective quantities. 
The operators $Q_m$ and $L_{m'}$ satisfy the commutation relation, 
\begin{eqnarray}
[Q_m,L_{m'}]=-\sqrt{6}(2m1m'|2m+m')Q_{m+m'}, 
\end{eqnarray}
which is identical to that for SU(3) by Elliott \cite{Elliott}. 
The commutation relation of $Q_m$ with $Q_{m'}$, however, involves not only $L_{m+m'}$ so as the case of SU(3) but also octupole operators $S_{m+m'}$.
Thus, the algebra of the 24 transformation generators $c^\dagger_{\mu i}c_{\mu'i}$ that appear in the Hamiltonian $H$ in Eq. (19) is extended to SU(5) \cite{Lichtenberg}, which employs, in addition to $L_m$ and $Q_m$, the octupole and hexadecapole operators, $S_m$ and $R_m$. 
The commutation relations of these operators for SU(5) algebra are expressed in terms of $6j$ symbols: If we use the unified expression for these multi-pole operators, 
\begin{eqnarray}
O_{lm}= \left \{ \begin{array}{ll}
L_m, \ \ \ \ \  & {\rm for} \ \ \  l=1, \\
Q_m, \ \ \ \ \  & {\rm for} \ \ \  l=2, \\
S_m, \ \ \ \ \  & {\rm for} \ \ \  l=3, \\
R_m, \ \ \ \ \  & {\rm for} \ \ \  l=4, 
\end{array} \right. 
\end{eqnarray}
the commutation relations are 
\begin{widetext}
\begin{eqnarray}
[O_{l_1m_1},O_{l_2m_2}]&=&\sum_{l(=l_1+l_2+{\rm odd})}(-1)^l2\sqrt{10(2l_1+1)(2l_2+1)} \left \{  \begin{array}{lll} 
2 & 2 & l \\
l_1 & l_2 & 2 
\end{array} \right \} \nonumber \\
&&\times(l_1m_1l_2m_2|lm_1+m_2)O_{lm_1+m_2}. 
\end{eqnarray}
\end{widetext}

The Casimir operator for SU(5) is defined as 
\begin{eqnarray}
C=L^\dagger\cdot L+Q^\dagger\cdot Q+S^\dagger\cdot S+R^\dagger\cdot R. 
\end{eqnarray}
The quadrupole-quadrupole interaction Hamiltonian $H$ between a pair of $d$ nucleons in Eq. (19) is expressed in terms of the quadrupole operators $Q_m$ as 
\begin{eqnarray}
H=-\kappa Q^\dagger\cdot Q=-\kappa \sum^2_{m=-2}Q^\dagger_mQ_m \label{3} 
\end{eqnarray}
and the interaction Hamiltonian $H_{sd}$ to excite $d$ nucleons to $s$ state and vice versa, 
\begin{eqnarray}
H_{sd}=-{\kappa'\over2\sqrt{6}}\sum_{mii'}(-1)^m\{c^\dagger_{mi}c^\dagger_{-mi'}c_{si'}c_{si}+{\rm H. c.}\}. 
\end{eqnarray}
The quadrupole operators conjugate to $Q_m$ are defined to be 
\begin{eqnarray}
Q^\dagger_m=(-1)^mQ_{-m}. 
\end{eqnarray}
The coupling constants $\kappa$ and $\kappa'$ of the
quadrupole-quadrupole interactions $H$ and $H_{sd}$ in Eq's. (31) and
(32) are determined in terms of the meson quadrupole mean field
$\bar{\varphi}_{2m}$ overlapped with the nucleon wave functions: 
\begin{widetext}
\begin{eqnarray}
\kappa&=&{g^2\over20\omega^2_2}\sum_\mu\int\int \psi^*_{2\mu-m}(\vec{r})\bar{\varphi}^*_{2m}(\vec{r})\psi_{2\mu}(\vec{r})\psi^*_{2-\mu+m}(\vec{r'})\bar{\varphi}_{2m}(\vec{r'})\psi_{2-\mu}(\vec{r'})d^3rd^3r', \nonumber \\
\kappa'&=&{\sqrt{6}g^2\over\omega^2_2}\int\int \psi^*_{2m}(\vec{r})\bar{\varphi}_{2m}(\vec{r})\psi_{s}(\vec{r})\psi^*_{2-m}(\vec{r'})\bar{\varphi}_{2-m}(\vec{r'})\psi_{s}(\vec{r'})d^3rd^3r'. 
\end{eqnarray}
\end{widetext}
The values of $\kappa$ and $\kappa'$ that are determined 
in the Hartree calculations are of the order of 0.1 MeV 
for the $s$-$d$ shell nuclei \cite{Kohmura2}. 
In the numerical calculation in Section IV, 
we assume that $\kappa=\kappa'$. 

The present interaction Hamiltonian $H=-\kappa Q^\dagger\cdot Q$ in Eq. (\ref{3}) commutes with the Casimir operator $C$,
\begin{eqnarray}
[H,C]=0. 
\end{eqnarray}
Therefore, eigenstates of the Hamiltonian $H$ 
are concurrently eigenstates of the Casimir operator $C$. 
Eigenstates of the Casimir operator $C$ with a same eigenvalue 
constitute a representation of SU(5). 
Any eigenstates of the interaction Hamiltonian $H$ 
can be classified by representations of SU(5) with a given eigenvalue 
of the Casimir operator $C$. 
Since the Hamiltonian $H$ is rotational symmetric, we obtain 
\begin{eqnarray}
[H, L_0]=0, \ \ \  [H, L_{\pm1}]=0, \ \ \  [H, \vec{L}^2]=0: 
\end{eqnarray} 
The eigenstates of the Hamiltonian $H$ are simultaneously 
eigenstates $|L, M\rangle$ of the orbital angular momentum operators 
${\vec{L}}^2$ and $L_0$. 
The eigenvalues of the Hamiltonian $H$ depend 
on the representation of SU(5) and on the angular momentum $L$ 
but not on $M$. 

In the previous paper \cite{Kohmura1}, 
applying the SU(5) Hamiltonian $H$ in Eq. (\ref{3}) 
to some typical systems of two and four $d$-state nucleons, 
we demonstrated that the Hamiltonian $H$ has many appropriate aspects 
for realistic systems, where we solved the eigenequations for $H$ and $C$, 
\begin{eqnarray}
H\Psi_{(k)}&=&\varepsilon_k\Psi_{(k)}, \\
C\Psi_{(k)}&=&\lambda^2_k\Psi_{(k)}. 
\end{eqnarray}
Antisymmetrized states of two $d$ nucleons with a unique combination 
of spin and isospin such as two protons with spin up, 
which have nuclear orbital angular momentum $L=1$ or 3, 
are classified into the 10-dimensional representation of SU(5). \\

\subsection{ Representations of SU(5) for nuclear states of ${}^{28}$Si}

The SU(5) Hamiltonian $H=-\kappa Q^\dagger\cdot Q$ in Eq. (31) 
yields the deformed Hartree states of an $s$-$d$ major-shell nucleus: 
The Hartree states of the nucleus ${}^{28}$Si, 
which has 12 valence nucleons in the $s$-$d$ major shell, 
are determined by the Hartree Hamiltonian for $H$, 
\begin{eqnarray}
H_{\rm Hart}=-2\kappa\langle Q_0\rangle Q_0+\kappa\langle Q_0\rangle^2. 
\end{eqnarray}

Since the Hartree Hamiltonian $H_{\rm Hart}$ is 
independent of the spin and isospin of nucleons, 
the nuclear wave functions for the nucleons in the $s$-$d$ major shell 
are expressed by a product of the four antisymmetrized wave functions, 
i.e., that for three protons with spin up, that 
for three protons with spin down, that for three neutrons 
with spin up and that for three neutrons with spin down. 

In the oblate Hartree state of the nucleus $^{28}$Si polarized 
along $z$-axis, the single-particle states $(d_2\ d_{-2}\ s)^4$ 
are occupied. 
On the whole basis set of nuclear oblate states 
in the nucleon configurations $d^8s^4$, 
which have 8 nucleons of the 12 valence nucleons in the $d$ 
states and 4 nucleons in the $s$ state, 
we diagonalize the Hamiltonian H in Eq. (31) to determine 
the nuclear oblate states with a given angular momentum $L$. 
The 12-nucleon system has 3 nucleons, i.e., two $d$ nucleons 
and one $s$ nucleon, for each a unique combination of spin 
and isospin of nucleons, such as protons with spin up, for example.  
The orbital wave functions of the three nucleons for a unique
combination of spin and isospin are antisymmetrized 
with the compound orbital angular momentum $L=1$ or 3. 
The oblate basis states for the 12-nucleon system 
are formulated by a product of the four antisymmetrized wave functions 
for each a set of three nucleons with a unique combination of spin 
and isospin. 
Since the antisymmetrized wave functions of two $d$ nucleons 
and one $s$ nucleon with a unique combination of spin and isospin 
are expressed by the 10-dimensional representation of SU(5) 
as mentioned at the end of Subsection A, the nuclear states 
of 12 nucleons in the nucleon configurations $d^8s^4$ 
are classified by the representations of SU(5) that 
are formed by a product $10\times10\times10\times10$ 
of SU(5) (see Table I(a)). 
The Hamiltonian $H=-\kappa Q^\dagger\cdot Q$ in Eq. (31) 
is diagonalized in the space of the nuclear basis states 
in the 490-dimensional representation formed by the product 
$10\times10\times10\times10$ for the nucleon configurations $d^8s^4$. 
The ground state of the Hamiltonian $H$ 
in the nucleon configurations $d^8 s^4$ 
is the lowest-lying state in the 490-dimensional representation. 
This state is identified as the ground state 
of oblate states with the nuclear angular momentum $L=0$. 
The 490-dimensional representation exhibits the rotational ground band 
in the space of the nucleon configurations $d^8s^4$ for the Hamiltonian $H$. 
The structure of the 490-dimensional representation 
is shown by the Young tableau in Fig. 1(a). 

On the other hand, in the prolate Hartree state polarized 
along $z$-axis, the single-particle states $(d_1\ d_{-1}\ d_0)^4$ 
are occupied: The nucleus has the 12 valence nucleons in the $d$ states. 
On the whole basis set of nuclear prolate states 
in the nucleon configurations $d^{12}$, 
which have all the 12 valence nucleons in the $d$ states, 
we diagonalize the Hamiltonian H in Eq. (31) to determine 
the nuclear prolate states with a given angular momentum $L$. 
The prolate basis states for the 12 nucleons in the $d$ states 
are expressed by a product of the four antisymmetrized 
orbital wave functions for three $d$ nucleons 
with a unique combination of spin and isospin: 
Nuclear prolate states of 12 nucleons in the nucleon 
configurations $d^{12}$ are classified by the representations 
of a product $10^*\times10^*\times10^*\times10^*$ 
of the $10^*$-dimensional representations for the antisymmetrized 
wave functions of three $d$ nucleons with a unique combination of spin 
and isospin. 
These SU(5) representations for the nucleon configurations $d^{12}$ 
are conjugate to the representations for the nucleon configurations $d^8s^4$. 
The ground state of the Hamiltonian $H$ 
in the configurations $d^{12}$ is the lowest-lying state 
in the 490*-dimensional representation formed by the product 
$10^*\times10^*\times10^*\times10^*$ of the $10^*$-dimensional 
representations. 
This state is identified as the ground state of prolate states 
with the nuclear angular momentum $L=0$. 
The energy spectrum for the Hamiltonian $H$ for the nucleon 
configurations $d^{12}$ in the space of the 490*-dimensional 
representation is of a same structure as the energy spectrum 
for the nucleon configurations $d^8s^4$ in the space 
of the 490-dimensional representation: The $490^*$-dimensional 
representation exhibits the rotational ground band 
in the space of the nucleon configurations $d^{12}$ 
for the Hamiltonian $H$ in Eq. (31). 
The structure of the $490^*$-dimensional representation 
is shown by the Young tableau in Fig. 1(b). 

Although the Hamiltonian $H$ in Eq. (31) does not couple 
the prolate states with the oblate states, the $s$-$d$ 
interaction $H_{sd}$ does in second-order processes. 
Therefore, we introduce the intermediate nuclear states 
in the nucleon configurations $d^{10}s^2$ 
which couple the 490-dimensional oblate states 
in the configurations $d^8s^4$ with the $490^*$-dimensional prolate
states in the configurations $d^{12}$. 
The significant nuclear states in the configurations $d^{10}s^2$ 
intermediate between the prolate and oblate states 
are in the representations of SU(5) that are formed 
by a product $10^*\times10^*\times10\times10$ 
of the 10- and $10^*$-dimensional representations 
for three nucleons in the configurations $d^2s$ and $d^3$, 
respectively, for a unique combination of spin and isospin (see Table I(b)). 
We diagonalize the Hamiltonian $H$ on the basis states 
in the nucleon configurations $d^{10}s^2$. 
The lowest-lying eigenenergy state of the Hamiltonian $H$ 
in the nucleon configurations $d^{10}s^2$ is classified 
into the 1176-dimensional representation. 
The 1176-dimensional representation exhibits the rotational ground band 
in the space of the nucleon configurations $d^{10}s^2$ 
for the Hamiltonian $H$. 
The structure of the 1176-dimensional representation 
is shown by the Young tableau in Fig. 1(c).  

\begin{figure}[h]
\begin{center}
\includegraphics[width=0.9\hsize]{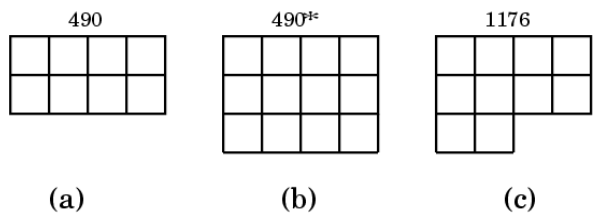}
\end{center}
\caption{The Young tableaus for the (a) 490-, (b) $490^*$- and (c) 1176-dimensional representations of SU(5). Each one box represents a $d$-state nucleon.}
\label{fig1}
\end{figure}

The eigenstates of $H$ are classified by SU(5) representations. 
In Fig.2, we show the eigenenergies of $H$ in Eq. (31) 
in the three representations, i.e., the (a) 490- and $490^*$-, 
and (b) 1176-dimensional representations. 
We see that the ground state band features a rotational structure 
of the nuclear energy levels in each a representation. 

\begin{figure}[h]
\begin{center}
\includegraphics[width=0.9\hsize]{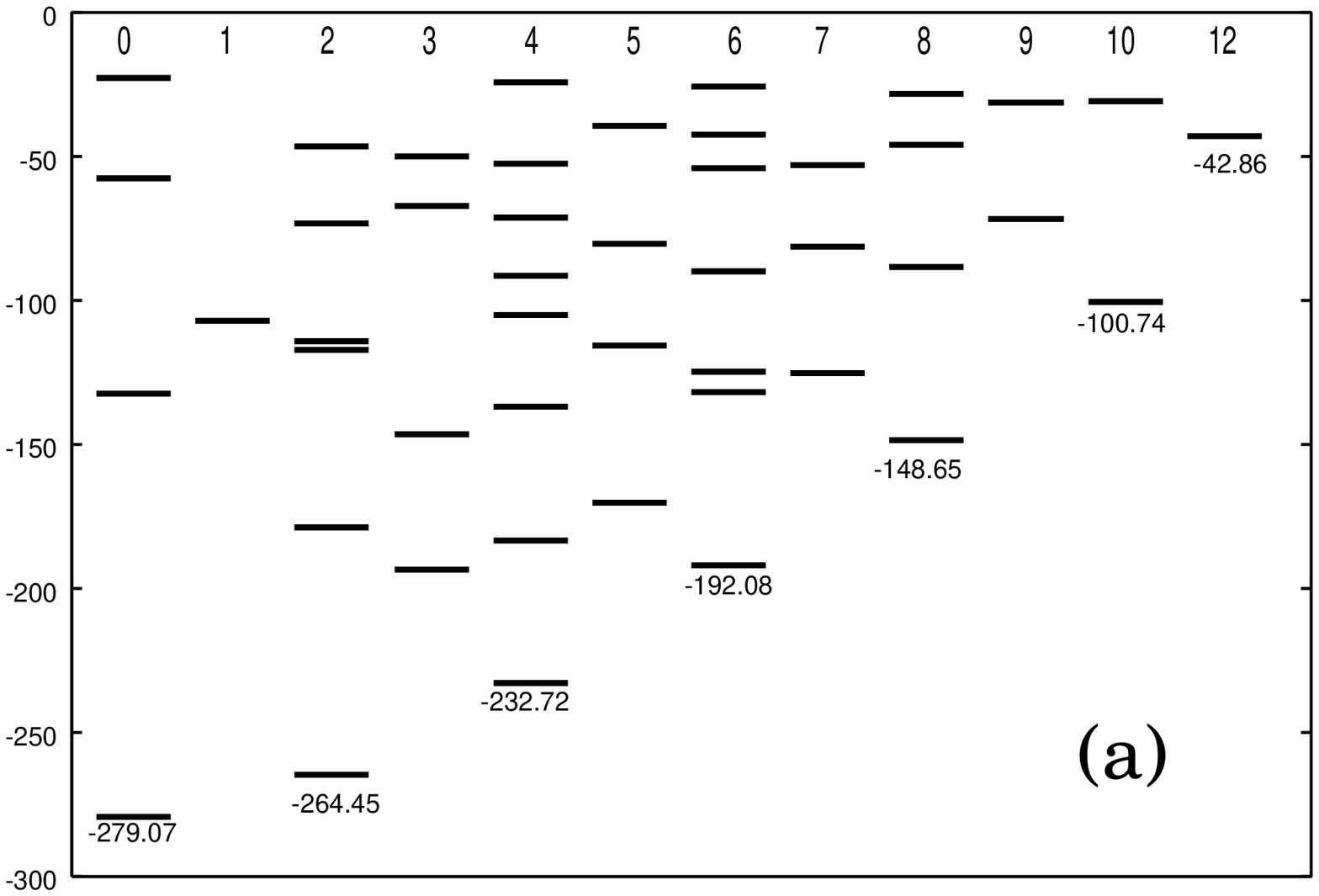}
\end{center}

\begin{center}
\includegraphics[width=0.9\hsize]{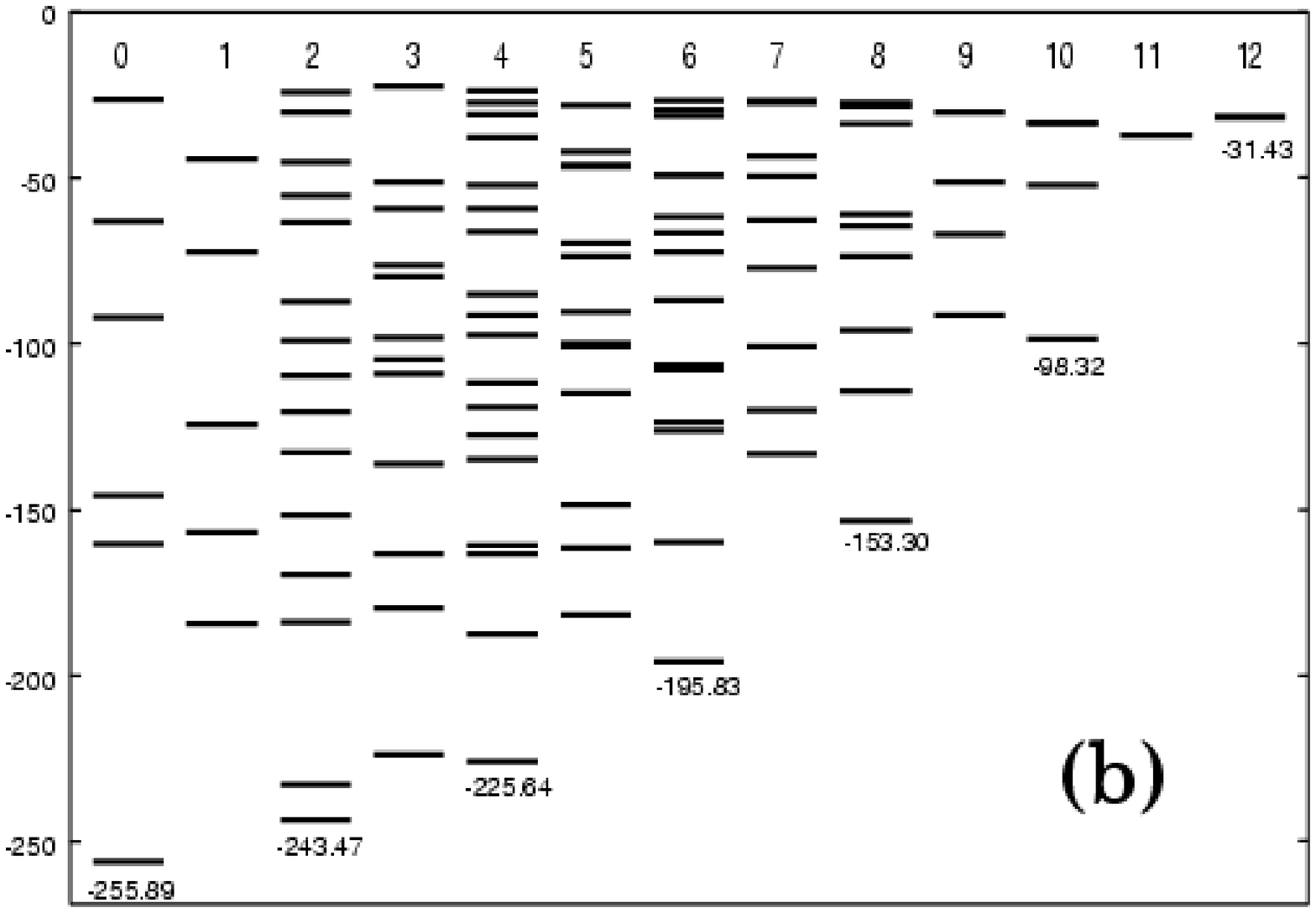}
\end{center}
\caption{The energy spectra for (a) the 490-dimensional 
representation and for (b) the 1176-dimensional representation. 
The energy spectrum for the $490^*$-dimensional representation 
is of a same structure as that for the 490-dimensional representation.  
The energies of the nuclear levels are shown in unit of $\kappa$. 
The integral numbers shown at the top of the figures stand 
for the angular momenta $L$ of the nuclear states.} 
\label{fig2b}
\end{figure}

\begin{table*}
\caption{The nuclear states of the system of 12 nucleons in the configurations (a) $d^8s^4$ and (b) $d^{10}s^2$ are classified by the representations of SU(5). 
The nuclear states of 12 nucleons in the configurations $d^{12}$ are classified into the representations conjugate to those for the configurations $d^8s^4$. 
The dimension and eigenvalue $\lambda^2$ of the Casimir operator $C$ for each a representation are shown. 
The column headed as Multiplicity shows the multiplicity of the representations with a same dimension and eigenvalue $\lambda^2$. 
Each a representation contains the nuclear states specified by their angular momenta $L$ with the multiplicities $M$, which are shown as $L^M$.} 
\label{I}

(a) The nuclear states in the nucleon configurations $d^8s^4$. 

\begin{tabular} {r|c|c|l}
Dimension & $\lambda^2$ & Multiplicity & \ \  $L^M$ \\ \hline
1050 \ \  & 352 & 3 & 11\ $10^2$\ $9^4$\ $8^6$\ $7^{10}$\ $6^{12}$\ $5^{15}$\ $4^{15}$\ $3^{16}$\ $2^{12}$\ $1^9$\ $0^2$ \\
560 \ \  & 312 & 2 & 10\ 9\ $8^4$\ $7^4$\ $6^8$\ $5^7$\ $4^{11}$\ $3^7$\ $2^{10}$\ $1^3$\ $0^4$ \\
490 \ \  & 432 & 1 & 12\ $10^2$\ $9^2$\ $8^4$\ $7^3$\ $6^7$\ $5^4$\ $4^8$\ $3^4$\ $2^6$\ 1\ $0^4$ \\
450 \ \  & 272 & 3 & 9\ $8^2$\ $7^4$\ $6^5$\ $5^8$\ $4^8$\ $3^9$\ $2^6$\ $1^6$\ 0 \\
315 \ \  & 272 & 3 & 9\ 8\ $7^3$\ $6^3$\ $5^6$\ $4^5$\ $3^7$\ $2^4$\ $1^5$ \\ 
210 \ \  & 232 & 6 & 8\ 7\ $6^3$\ $5^3$\ $4^5$\ $3^4$\ $2^5$\ $1^2$\ $0^2$ \\
175 \ \  & 192 & 7 & 7\ $6^2$\ $5^3$\ $4^4$\ $3^5$\ $2^4$\ $1^3$\ 0 \\
40 \ \  & 132 & 8 & 5\ 4\ 3\ $2^2$\ 1 \\
35 \ \  & 192 & 1 & 6\ 4\ 3\ 2\ 0 \\
15 \ \  & 112 & 3 & 4\ 2\ 0 \\
10 \ \  & 72 & 6 & 3\ 1 
\end{tabular}


\quad \\
\quad \\
(b) The nuclear states in the nucleon configurations $d^{10}s^2$.

\begin{tabular} {r|c|c|l}
Dimension & $\lambda^2$ & Multiplicity & \ \  $L^M$ \\ \hline
176 \ \  & 400 & 1 & 12\ 11\ $10^4$\ $9^4$\ $8^9$\ $7^9$\ $6^{15}$\ $5^{13}$\ $4^{18}$\ $3^{12}$\ $2^{15}$\ $1^5$\ $0^6$ \\
1024 \ \  & 300 & 4 & 10\ $9^3$\ $8^5$\ $7^9$\ $6^{13}$\ $5^{15}$\ $4^{18}$\ $3^{18}$\ $2^{14}$\ $1^{10}$\ $0^4$ \\
700 \ \  & 360 & 1 & 11\ 10\ $9^3$\ $8^4$\ $7^7$\ $6^7$\ $5^{11}$\ $4^9$\ $3^{11}$\ $2^7$\ $1^7$ \\
$700^*$ \ \  & 360 & 1 & 11\ 10\ $9^3$\ $8^4$\ $7^7$\ $6^7$\ $5^{11}$\ $4^9$\ $3^{11}$\ $2^7$\ $1^7$ \\
200 \ \  & 240 & 2 & 8\ 7\ $6^3$\ $5^3$\ $4^5$\ $3^3$\ $2^5$\ 1\ $0^2$ \\
175 \ \  & 240 & 3 & 8\ 7\ $6^2$\ $5^3$\ $4^4$\ $3^3$\ $2^4$\ $1^2$\ 0 \\ 
$175^*$ \ \  & 240 & 3 & 8\ 7\ $6^2$\ $5^3$\ $4^4$\ $3^3$\ $2^4$\ $1^2$\ 0 \\ 
126 \ \  & 200 & 4 & 7\ 6\ $5^3$\ $4^2$\ $3^4$\ $2^2$\ $1^3$ \\
$126^*$ \ \  & 200 & 4 & 7\ 6\ $5^3$\ $4^2$\ $3^4$\ $2^2$\ $1^3$ \\
75 \ \  & 160 & 9 & 6\ 5\ $4^2$\ $3^2$\ $2^3$\ 1\ 0 \\
24 \ \  & 100 & 8 & 4\ 3\ 2\ 1 \\
1 \ \  & 0 & 3 & 0 
\end{tabular}
\end{table*}

\section{ COLLECTIVE TUNNELING TRANSITIONS BETWEEN DEFORMED HARTREE STATES} 

The $s$-$d$ interaction Hamiltonian $H_{sd}$ couples 
a nuclear oblate state in the 490-dimensional representation 
with a prolate state in the $490^*$-dimensional representation 
through an intermediate state in the 1176-dimensional representation 
as is shown in Fig. 3: Nuclear tunneling transitions 
from an oblate state to a prolate state proceed 
through an intermediate state, hopping in a two-step process \cite{Arve}.  
Taking into account the Hamiltonian $H_{sd}$, 
we solve the secular equation for the nuclear Hamiltonian 
$H_{\rm nucl}$ in Eq. (15), 
\begin{eqnarray}
H_{\rm nucl}\Psi_k=E_k\Psi_k, 
\end{eqnarray}
in the combined space of the nuclear states 
in the 490-, 1176- and 490*-dimensional representations of SU(5). 
The eigenstates of the Hamiltonian $H_{\rm nucl}$, 
which commutes with the angular momentum operators $L_m$, 
are simultaneously eigenstates of the angular momentum. 
Since the oblate Hartree states in the configurations $d^8s^4$ 
are degenerate one to one correspondingly to the prolate Hartree states 
in the configurations $d^{12}$, there appear two almost degenerate 
lowest-lying nuclear states for each of angular momenta $L=0,\ 2$ and 4. 
The calculated energy splitting $\Delta E$ 
between the almost degenerate pair of the ground state 
and the first excited state for $L$=0, 2 and 4 are shown in Table II. 
We obtain approximate values $\Delta E$=1.37, 0.94 and 1.00 
of the energy splitting in unit of $\kappa$ for $L=0,\ 2$ and 4 
in the simplified calculations which use only one lowest-lying state 
for each of the three sets of configurations $d^8s^4,\ d^{10}s^2$ 
and $d^{12}$ \cite{kohmura3}. 
The perturbational expression for the energy splitting 
for the three-state system, 
\begin{eqnarray}
\Delta E={2h^2\over\varepsilon^{1176}_0-\varepsilon^{490}_0}, 
\end{eqnarray}
yields a reasonablly approximate value. 
In the above expression, $\varepsilon^{490}_0$ 
and $\varepsilon^{1176}_0$ stand for the energies 
of the ground states $\Psi^{490}_{(0)}$ and $\Psi^{1176}_{(0)}$ for given $L$ in the 490- ($490^*$-) 
and 1176-dimensional representation spaces, respectively, 
and $h$ is the matrix element of the interaction Hamiltonian 
$H_{sd}$ between the ground states of the Hamiltonian $H$: 
\begin{eqnarray}
h=\langle\Psi^{1176}_{(0)}|H_{sd}|\Psi^{490}_{(0)}\rangle. \nonumber 
\end{eqnarray}

\begin{figure}[h]
\begin{center}
\includegraphics[width=0.7\hsize]{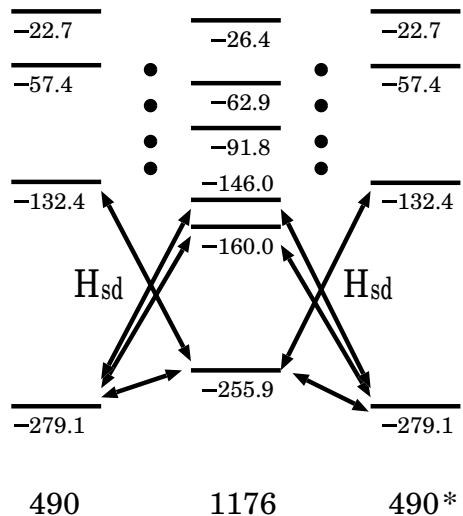}
\end{center}
\caption{The processes of the tunneling transitions 
between the nuclear oblate states in the 490-dimensional representation 
and the nuclear prolate states in the $490^*$-dimensional 
representation with angular momentum $L=0$ are illustrated. 
The tunneling transitions proceed through nuclear intermediate states 
in the 1176-dimensional representation in terms of the Hamiltonian $H_{sd}$.} 
\label{fig3}
\end{figure}

\begin{table}
\caption{The energies $E_k$ of the nuclear ground state and first excited state for angular momentum $L=0,\ 2$ and 4 and their splitting $\Delta E$ calculated in Eq. (40) are shown in unit of $\kappa$.}
\label{II}
\begin{center}
\begin{tabular} 
{r|c|c|r}
\hline 
$L$ & Ground State & First Excited State & $\Delta$E \\ \hline 
0 & -280.47 & -279.14 & 1.33 \\
2 & -265.46 & -264.71 & 0.75 \\
4 & -233.87 & -232.94 & 0.93 \\ \hline 
\end{tabular}
\end{center}
\end{table}

Now, we formulate the collective tunneling transitions 
between prolate and oblate Hartree states. 
In a real-time description, the probability of the collective 
tunneling transitions from an initial state $\Psi_i$ 
to a final state $\Psi_f$ is expressed in terms of the 
eigenstates $\Psi_k$ and eigenenergies $E_k$ 
of the nuclear Hamiltonian $H_{\rm nucl}$ determined in Eq. (40) as  
\begin{eqnarray}
P_{fi}(t)&=&|\langle\Psi_f|e^{-iH_{\rm nucl}t}|\Psi_i\rangle|^2 \nonumber \\  
&=&|\sum_k\langle\Psi_f|\Psi_k\rangle e^{-iE_kt}\langle\Psi_k|\Psi_i\rangle|^2. 
\end{eqnarray}

We have performed two kinds of the calculations. 
First, we calculated the probabilities of the nuclear collective 
tunneling transitions for given orbital angular momentum $L$ 
from the prolate ground state in the 490*-dimensional 
representation space for the configurations $d^{12}$ 
to the oblate ground state in the 490-dimensional 
representation space for the configurations $d^8s^4$. 
Since the Hamiltonian $H_{\rm nucl}$ is rotationally 
symmetric, the transitions between the two ground states 
conserve the orbital angular momentum $L$. 
The calculated transition probabilities between the nuclear ground
states for the orbital angular momentum $L=0$, 2 and 4 
are shown in Fig. 4. 
We see that each of the transition probabilities features 
a gross structure of the harmonic collective tunneling oscillations 
perturbed by small quantum mechanical fluctuations.  

\begin{figure}[h]
\begin{center}
\includegraphics[width=1.0\hsize]{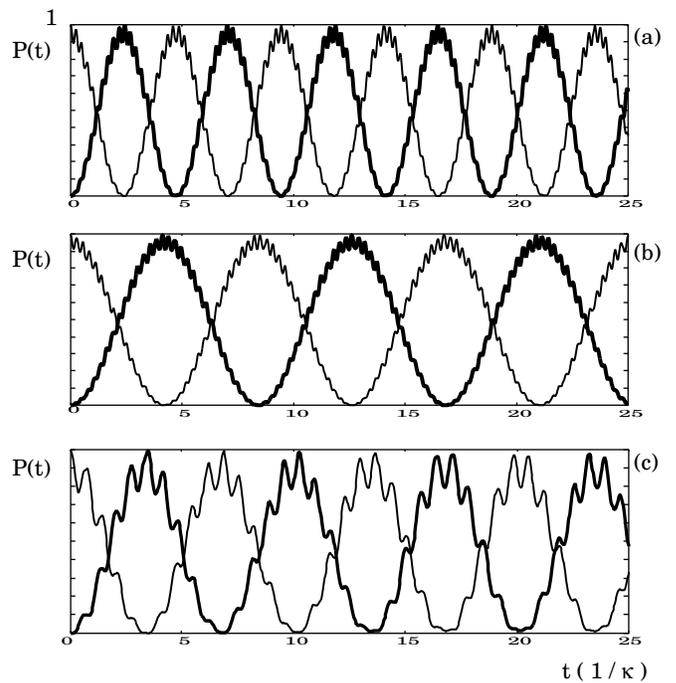}
\end{center}
\caption{The calculated probability of the tunneling transitions from the prolate ground state to the oblate ground state of the nucleus with angular momentum (a) $L=0$, (b) 2 and (c) 4. The probability of the return transitions to the initial prolate ground state is also shown in Figures (a) to (c). The transition probabilities feature grossly the harmonic oscillation of the nucleus between the prolate and oblate states. The frequency of the harmonic oscillation is determined by the energy splitting $\Delta E$ between the ground state and the first excited state of the nuclear Hamiltonian $H_{\rm nucl}$ for given angular momentum $L$.}
\label{fig4}
\end{figure}

Secondly, we calculated the probabilities 
of the tunneling transitions from a prolate Hartree state 
to an oblate Hartree state. 
Hereafter, the prolate Hartree state polarized 
along $z$-axis is called $z$-prolate state. 
We calculate the probability in the two cases of the tunneling 
transitions starting with the $z$-prolate state 
to a final oblate Hartree state. 
One is the case of the tunneling to the oblate Hartree state 
polarized along $z$-axis ($z$-oblate state) and the other 
is to a linear combination of the two oblate Hartree states, 
i.e., the oblate state polarized along $x$-axis ($x$-oblate state) 
and the oblate state polarized along $y$-axis ($y$-oblate state),
\begin{eqnarray}
|\Psi_f\rangle={1\over\sqrt{2}}(|x{\rm -oblate}\rangle+|y{\rm -oblate}\rangle). 
\end{eqnarray}
We show the calculated transition probabilities in Fig. 5. 
We see that the transition probability from the $z$-prolate state 
to the $x$- and $y$-oblate states is suppressed by about 40 percent, 
compared with the transition probability from the $z$-prolate state 
to the $z$-oblate state. \\

\begin{figure}[h]
\begin{center}
\includegraphics[width=1.0\hsize]{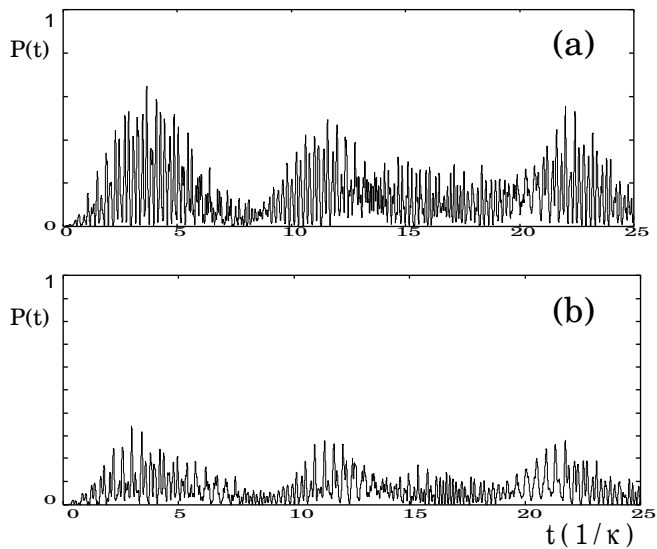}
\end{center}
\caption{The calculated probabilities of the tunneling transitions from a prolate Hartree state to an oblate Hartree state. We show the probabilities of the transitions from $z$-prolate Hartree state to (a) $z$-oblate Hartree state and to (b) the linear combination of $x$- and $y$-oblate states shown in Eq. (43).}
\label{fig5}
\end{figure}

\section{ NUCLEAR TUNNELING TRANSITIONS 
THROUGH TRI-AXIALLY ASYMMETRIC DEFORMED STATES}

We have formulated the nuclear collective tunneling transitions 
between a prolate Hartree state and an oblate Hartree state, 
taking into account the angular momentum conservation. 
The deformed Hartree states may be polarized along any axis. 
We may consider any ``twisted'' tunneling transitions, 
such as that from the $z$-prolate state to the $x$-oblate state. 

In the course of a tunneling transition from $z$-prolate state 
to $z$-oblate state, if the nucleus keeps rotational symmetry 
around $z$-axis in the shape, the nucleus proceeds 
through a spherical intermediate state with quadrupole moment 
$\langle-Q_0\rangle=0$ lying at the top of a high potential energy barrier. 
The high potential energy barrier on the way suppresses 
the transition probability. 
We will discuss in this Section that 
in the twisted collective tunneling transitions, 
such as the transitions from $z$-prolate state either 
to $x$-oblate state or to $y$-oblate state, 
the nucleus proceeds through tri-axially asymmetric deformed 
intermediate states \cite{Bohr}\cite{Ring}, i.e., 
through a lower potential energy barrier than the spherical 
potential energy barrier in the course of the direct tunneling 
transition from $z$-prolate state to $z$-oblate state. 

Hereafter we express the $z$-prolate and $z$-oblate Hartree states with spin and isospin $S=T=0$ explicitly in terms of occupied single-particle states as 
\begin{eqnarray}
|z{\rm -prolate}\rangle&=&|(d_1\ d_{-1}\ d_0)^4\rangle, \\
|z{\rm -oblate}\rangle&=&|(d_2\ d_{-2}\ s)^4\rangle. 
\end{eqnarray}
Rotating the above $z$-prolate and $z$-oblate states 
by 90 degrees around $y$-axis, we obtain the $x$-prolate 
and $x$-oblate states, respectively, which are expressed 
in terms of occupied single-particle states as  
\begin{widetext}
\begin{eqnarray}
|x{\rm -prolate}\rangle&=&|\{{1\over\sqrt{2}}(d_2-d_{-2})\ {1\over\sqrt{2}}(d_1-d_{-1})\ (-{1\over2}d_0+{\sqrt{3}\over2\sqrt{2}}(d_2+d_{-2}))\}^4\rangle, \nonumber \\
 \\
|x{\rm -oblate}\rangle&=&|\{{1\over\sqrt{2}}(d_1+d_{-1})\ ({\sqrt{3}\over2}d_0+{1\over2\sqrt{2}}(d_2+d_{-2}))\ s\}^4\rangle. 
\end{eqnarray}
\end{widetext}
We also obtain the $y$-prolate and $y$-oblate states, 
rotating the $z$-prolate and $z$-oblate states, 
respectively, by $-$90 degrees around $x$-axis, 
\begin{widetext}
\begin{eqnarray}
|y{\rm -prolate}\rangle&=&|\{{1\over\sqrt{2}}(d_2-d_{-2})\ {1\over\sqrt{2}}(d_1+d_{-1})\ ({1\over2}d_0+{\sqrt{3}\over2\sqrt{2}}(d_2+d_{-2}))\}^4\rangle, \nonumber \\ 
 \\
|y{\rm -oblate}\rangle&=&|\{{1\over\sqrt{2}}(d_1-d_{-1})\ (-{\sqrt{3}\over2}d_0+{1\over2\sqrt{2}}(d_2+d_{-2}))\ s\}^4\rangle. 
\end{eqnarray}
\end{widetext}
Note that the Hartree energies of these six deformed states are degenerate. 

From the above nucleon configurations (44)-(49) 
of Hartree states of the $s$ and $d$ major-shell nucleus, 
we see that the direct tunneling transitions 
from $z$-prolate state to $z$-oblate state take place, 
only in the case that all of the 12 nucleons 
in $d_1$, $d_{-1}$ and $d_0$ states are excited 
to $d_2$, $d_{-2}$ and $s$ states. 
The twisted tunneling transitions can, however, 
take place by exciting only four nucleons 
in a second order process of the $s$-$d$ interaction $H_{sd}$. 
The twisted tunneling transitions from $z$-prolate state 
to $x$-oblate state, for example, can take place 
by exciting four nucleons in ${1\over\sqrt{2}}(d_1-d_{-1})$ 
state to $s$ state in a second order process 
of the $s$-$d$ interaction $H_{sd}$, where the nucleons 
in the states ${1\over\sqrt{2}}(d_1+d_{-1})$ and $d_0$ are spectators, 
which are not affected by the interaction Hamiltonian $H_{sd}$. 

The second-order processes in $H_{sd}$ 
in the nuclear collective tunneling transitions 
from $z$-prolate state to $x$-oblate state proceed 
through intermediate states. 
One of the typical intermediate states for the tunneling 
transitions between the two deformed Hartree states 
is obtained by a square root of a product of the two Hartree states,
\begin{eqnarray}
&&|\{(z{\rm -prol})(x{\rm -obl})\}^{1\over2}\rangle=|\{{1\over\sqrt{2}}(d_1+d_{-1})\}^4 \nonumber \\
&&\{d_0\ ({\sqrt{3}\over2}d_0+{1\over{2\sqrt{2}}}(d_2+d_{-2}))\}^2\ \{{1\over\sqrt{2}}
    (d_1-d_{-1})\ s\}^2\rangle, \nonumber \\
&&
\end{eqnarray}
where the 4 nucleons expressed by each a pair of the parentheses $\{$ and $\}$ constitute a spatial symmetric state with spin and isospin $S=T=0$. 
This state is a tri-axially asymmetric deformed state with $\gamma$=30 where $\gamma$ is defined \cite{Bohr} as 
\begin{eqnarray}
{\rm tan}\gamma=-{\langle Q_2\rangle+\langle Q_{-2}\rangle\over\sqrt{2}\langle Q_0\rangle}. 
\end{eqnarray}
Note that we define the angle $\gamma$ 
along the direction opposite to that in Ref. \cite{Bohr}. 
The transition matrix elements for the second-order tunneling process 
from $z$-prolate state to $x$-oblate state 
through this intermediate state are  
\begin{eqnarray}
\langle\{(z{\rm -prol})(x{\rm -obl})\}^{1\over2}|H_{sd}|z{\rm -prolate}\rangle 
   &=& -{9\over2}\sqrt{8\over219}\kappa', \nonumber \\
   & &                                              \\
\langle x{\rm -oblate}|H_{sd}|\{(z{\rm -prol})(x{\rm -obl})\}^{1\over2}\rangle
   &=& -{9\over2}\sqrt{8\over219}\kappa'.  \nonumber \\
   & &                                               
\end{eqnarray}

Similarly the second-order processes in $H_{sd}$ 
in the nuclear collective tunneling transitions from $z$-prolate state 
to $y$-oblate state proceed through intermediate states. 
One of the typical intermediate states for the tunneling transitions 
between the two deformed Hartree states is obtained by a square root 
of a product of the two Hartree states, 
\begin{eqnarray}
&&|\{(z{\rm -prol})(y{\rm -obl})\}^{1\over2}\rangle=|\{{1\over\sqrt{2}}(d_1-d_{-1})\}^4 \nonumber \\ 
&&\{d_0\ (-{\sqrt{3}\over2}d_0+{1\over2\sqrt{2}}
     (d_2+d_{-2}))\}^2\ \{{1\over\sqrt{2}}(d_1+d_{-1})\ s\}^2\rangle, \nonumber \\
&&      
\end{eqnarray}
where the 4 nucleons expressed by each a pair of the parentheses $\{$ 
and $\}$ constitute a spatial symmetric state 
with spin and isospin $S=T=0$. 
This state is a tri-axially asymmetric deformed state with $\gamma=-30$. 
The transition matrix elements for the second-order tunneling process 
from $z$-prolate state to $y$-oblate state 
through this intermediate state are 
\begin{eqnarray}
\langle\{(z{\rm -prol})(y{\rm -obl})\}^{1\over2}|H_{sd}|z{\rm -prolate}\rangle
   &=& {9\over2}\sqrt{8\over219}\kappa', \nonumber \\
   & &                                             \\               
\langle y{\rm -oblate}|H_{sd}|\{(z{\rm -prol})(y{\rm -obl})\}^{1\over2}\rangle
   &=&{9\over2}\sqrt{8\over219}\kappa'. \nonumber \\
   & &                                            
\end{eqnarray}
The six Hartree states and six tri-axially asymmetric 
deformed intermediate states are schematically plotted 
in the $\beta$-$\gamma$ plane in Fig. 6. 

\begin{figure}[h]
\begin{center}
\includegraphics[width=1.0\hsize]{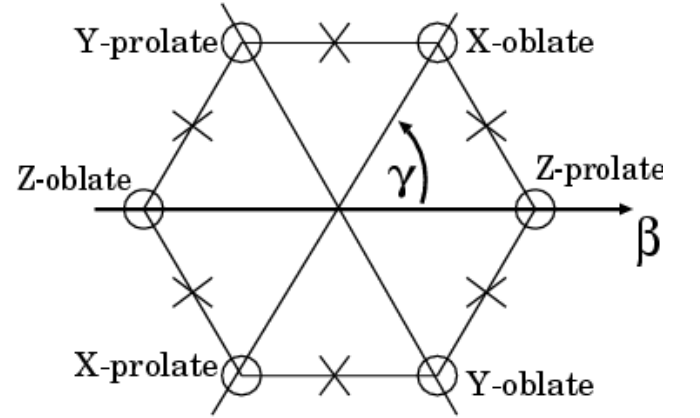}
\end{center}
\caption{The prolate Hartree states and oblate Hartree states polarized along $z$-, $x$- and $y$-axis are plotted with a circle and the intermediate states such as $|\{(x{\rm -obl})(y{\rm -prol})\}^{1\over2}\rangle$ in the twisted tunneling transitions are also plotted with a cross in the $\beta$-$\gamma$ plane.} 
\label{fig6}
\end{figure}

While the spherical intermediate states in the tunneling transitions 
of the nucleus ${}^{28}$Si are highly excited, 
the above intermediate states of the twisted 
collective tunneling transitions, 
which are tri-axially asymmetric deformed states, 
are not so highly excited. 
The twisted tunneling transitions 
through a lower potential energy barrier 
are not so much hindered as the direct tunneling transitions 
from $z$-prolate state to $z$-oblate state 
through a spherical intermediate state. 
Therefore we see that the tunneling transitions 
from $z$-prolate state to $z$-oblate state 
may take place mainly proceeding through some tri-axially 
asymmetric deformed intermediate states by several steps 
avoiding the direct tunneling through a spherical intermediate state: 
\begin{widetext}
\begin{eqnarray}
&&|z{\rm -prolate}\rangle \ \rightarrow \nonumber \\
&&\left \{ \begin{array} {ccc} 
|\{(z{\rm -prol})(y{\rm -obl})\}^{1\over2}\rangle & |\{(y{\rm -obl})(x{\rm -prol})\}^{1\over2}\rangle & |\{(x{\rm -prol})(z{\rm -obl})\}^{1\over2}\rangle \\
|\{(z{\rm -prol})(x{\rm -obl})\}^{1\over2}\rangle & |\{(x{\rm -obl})(y{\rm -prol})\}^{1\over2}\rangle & |\{(y{\rm -prol})(z{\rm -obl})\}^{1\over2}\rangle 
\end{array} \right \} \nonumber \\
&&\rightarrow \ |z{\rm -oblate}\rangle. 
\end{eqnarray}
\end{widetext}
The tunneling transitions through tri-axially asymmetric deformed intermediate states lying on a lower potential energy barrier contribute to enhance the probability of the tunneling transitions from $z$-prolate state to $z$-oblate state.
The tri-axially asymmetric deformed states 
such as $|\{(z{\rm -prol})(y{\rm -obl})\}^{1\over2}\rangle$ 
belong to the 1176-dimensional representation of SU(5). 
They are main components of the rotational ground band states  
in the representation and are also main components 
of $H_{sd}|z{\rm -prolate}\rangle$, as is shown in Table III. 
The numerical result reported in Section IV 
for the enhanced probability of the tunneling transitions 
from $z$-prolate state to $z$-oblate state reflects 
the significant contributions of various twisted 
tunneling processes through tri-axially asymmetric 
deformed intermediate states shown in Eq. (57). 

\begin{table}
\caption{The coefficients $c^n_{LM}$ of the normalized states of $H_{sd}|z{\rm -oblate}\rangle$ and of $H_{sd}|z{\rm -prolate}\rangle$ expanded in terms of the eigenstates of the Hamiltonian $H$ with angular momentum $L$ and $M$, and principal quantum number $n$ in the 1176-dimensional representation. The presently concerned values of the coefficients $c^n_{LM}$ only for $n=1$ and $M=0$ are shown. The double signs $\pm$ of the coefficient values correspond to the $|z{\rm -prolate}\rangle$ and $|z{\rm -oblate}\rangle$ states to be operated by $H_{sd}$, respectively.} 
\label{III}
\begin{center}
\begin{tabular} {r|r}
\hline 
$L\ M$ & $c^1_{L0}$ \\ \hline
0$\ $0 & 0.32 \\
2$\ $0 & $\pm$0.52 \\
4$\ $0 & -0.41 \\
6$\ $0 & $\pm$0.29 \\
8$\ $0 & 0.17 \\
10$\ $0 & $\pm$0.07 \\
12$\ $0 & -0.02 \\ \hline
\end{tabular}
\end{center}
\end{table}

\begin{table}
\caption{The coefficients $c^n_{LM}$ of the prolate and oblate Hartree states in Eq's. (44)-(49) expanded in terms of the eigenstates of the Hamiltonian $H$ with angular momentum $L$ and $M$, and principal quantum number $n$ in the 490- ($490^*$-) dimensional representation. The presently concerned values of the coefficients $c^n_{LM}$ only for $n=1$ and $M=0$ are shown. The summed probabilities $\sum_L(c^1_{L0})^2$ for the Hartree states are also shown.}
\label{IV}
\begin{center}
\begin{tabular} {c|c|c}
\hline
&  & $|x{\rm -oblate}\rangle$ \\ 
$L\ M$ & $|z{\rm -oblate}\rangle$ & $|y{\rm -oblate}\rangle$ \\
& $|z{\rm -prolate}\rangle$  & $|x{\rm -prolate}\rangle$ \\
&  & $|y{\rm -prolate}\rangle$ \\ \hline
0$\ $0 & 0.34 & 0.34  \\
2$\ $0 & -0.64 & 0.32 \\
4$\ $0 & 0.56 & 0.21 \\
6$\ $0 & 0.32 & -0.10 \\
8$\ $0 & 0.13 & 0.04 \\
10$\ $0 & 0.04 & -0.01 \\
12$\ $0 & 0.01 & 0.002 \\ \hline 
$\sum_L(c^1_{L0})^2$ & 0.96 & 0.27 \\ \hline 
\end{tabular}
\end{center}
\end{table}

The Hartree states defined in Eq's. (44)-(49) 
are expanded in terms of the eigenstates 
of the Hamiltonian $H$ with angular momentum $L$ 
and $M$, and principal quantum number $n$ 
in the 490- ($490^*$-) dimensional representation so as 
\begin{eqnarray}
|z{\rm -prolate}\rangle&=&\sum_{nL}c^n_{L0}|L,M=0,n\rangle, \\
|z{\rm -oblate}\rangle&=&\sum_{nL}c^n_{L0}|L,M=0,n\rangle. 
\end{eqnarray}
The calculated coefficients $c^1_{L0}$ are shown in Table IV.
While the Hartree states polarized along $z$-axis 
in Eq's. (44) and (45) are composed of only $M=0$ components 
of the angular momentum $L$, the Hartree states polarized either 
along $x$-axis or along $y$-axis in Eq's. (46)-(49) have $M\neq0$ 
components, with comparatively small $M=0$ components as is shown in Table IV. 
Since the angular momentum of the nucleus is conserved 
in the course of the tunneling transitions, the probability 
of the twisted tunneling transitions from $z$-prolate state 
to the combination of $x$- and $y$-oblate states in Eq. (43) 
as the final state is suppressed by about 40 percent compared 
with that of the tunneling transitions from $z$ prolate state 
to $z$-oblate state, as is seen in the numerical result in Section IV. \\ 

\section{DISCUSSIONS AND CONCLUSION} 

Using the physical quantities obtained in the Hartree calculations 
for the deformed Hartree states, 
we determine the Hamiltonian effective not only 
for the prolate and oblate Hartree states but also 
for the collective tunneling transitions between the two Hartree states. 
In the case of the nucleus of ${}^{28}$Si, 
which has a prolate Hartree state and an oblate Hartree state 
symmetrically with respect to the deformation parameter, 
the nuclear field Hamiltonian $H_{\rm nucl}$ in Eq. (1) yields 
the Hamiltonian $H$ for the interaction between $d$ nucleons 
and the Hamiltonian $H_{sd}$ to excite $d$ nucleons to $s$ state 
and vice versa. 
The interaction Hamiltonian $H$ between $d$ nucleons, 
which we have expressed in terms of the SU(5) algebra, 
yields the nuclear deformed Hartree states. 
The Hamiltonian $H_{sd}$ instead gives rise 
to the collective tunneling transitions between prolate states 
and oblate states. 
We have calculated the probabilities of the tunneling transitions 
between a prolate state and an oblate state in the real-time description. 

The spherically symmetric nuclear field Hamiltonian $H_{\rm nucl}$ 
in Eq. (1) yields not only the interaction Hamiltonian 
$H$ between $d$ nucleons in Eq. (19)
 but also the residual $s$-$d$ interaction $H_{sd}$ 
in Eq. (20) for a deformed $s$ and $d$ shell nucleus 
in the present theory beyond the Hartree approximation. 
The rotationally symmetric Hamiltonian $H$ in Eq. (19) 
recovers the rotational symmetry of the nuclear system 
by giving rise to the rotation of the deformed nucleus. 
Furthermore, the spherically symmetric nuclear field Hamiltonian 
$H_{\rm nucl}$ restores the spherical symmetry broken 
in a nuclear deformed Hartree state so that it gives rise to 
the nuclear collective tunneling transitions between deformed Hartree
states: The residual interaction Hamiltonian $H_{sd}$, 
acting as the restoring force for the spherical symmetry 
of the nuclear system to be recovered, gives rise to the collective 
tunneling transitions between prolate Hartree states 
and oblate Hartree states. 
Since the spherical intermediate states are highly excited 
lying at the top of the energy surface of the nucleus ${}^{28}$Si, 
the spherical symmetry broken in a prolate Hartree state 
is restored so that the nucleus tunnels to an oblate Hartree state,
 proceeding through tri-axially asymmetric deformed states. 
The ground state of the nuclear Hamiltonian $H_{\rm nucl}$ 
in Eq. (15) contains not only the Hartree states 
in Eq's. (44)-(49) but also the tri-axially asymmetric deformed 
states such as in Eq's. (50) and (54). \\


\begin{thebibliography}{14}
\expandafter\ifx\csname natexlab\endcsname\relax\def\natexlab#1{#1}\fi
\expandafter\ifx\csname bibnamefont\endcsname\relax
  \def\bibnamefont#1{#1}\fi
\expandafter\ifx\csname bibfnamefont\endcsname\relax
  \def\bibfnamefont#1{#1}\fi
\expandafter\ifx\csname citenamefont\endcsname\relax
  \def\citenamefont#1{#1}\fi
\expandafter\ifx\csname url\endcsname\relax
  \def\url#1{\texttt{#1}}\fi
\expandafter\ifx\csname urlprefix\endcsname\relax\def\urlprefix{URL }\fi
\providecommand{\bibinfo}[2]{#2}
\providecommand{\eprint}[2][]{\url{#2}}

\bibitem[{\citenamefont{Kohmura}(2001)}]{Kohmura}
\bibinfo{author}{\bibfnamefont{T.}~\bibnamefont{Kohmura}},
  \bibinfo{journal}{Prog.\ Theor.\ Phys.} \textbf{\bibinfo{volume}{106}},
  \bibinfo{pages}{471} (\bibinfo{year}{2001}).

\bibitem[{\citenamefont{Kohmura et~al.}(2003)\citenamefont{Kohmura, Maruyama,
  and Y.Hashimoto}}]{Kohmura1}
\bibinfo{author}{\bibfnamefont{T.}~\bibnamefont{Kohmura}},
  \bibinfo{author}{\bibfnamefont{M.}~\bibnamefont{Maruyama}}, \bibnamefont{and}
  \bibinfo{author}{\bibnamefont{Y.Hashimoto}}, \bibinfo{journal}{Phys.\ Rev.}
  \textbf{\bibinfo{volume}{C67}}, \bibinfo{pages}{054316}
  (\bibinfo{year}{2003}).

\bibitem[{\citenamefont{Bohr and Mottelson}(1975)}]{Bohr}
\bibinfo{author}{\bibfnamefont{A.}~\bibnamefont{Bohr}} \bibnamefont{and}
  \bibinfo{author}{\bibfnamefont{B.~R.} \bibnamefont{Mottelson}},
  \emph{\bibinfo{title}{Nuclear Structure Vol.II}}
  (\bibinfo{publisher}{Benjamin}, \bibinfo{year}{1975}).

\bibitem[{\citenamefont{Ring and Schuck}(1980)}]{Ring}
\bibinfo{author}{\bibfnamefont{P.}~\bibnamefont{Ring}} \bibnamefont{and}
  \bibinfo{author}{\bibfnamefont{P.}~\bibnamefont{Schuck}},
  \emph{\bibinfo{title}{The Nuclear Many-Body Problem}}
  (\bibinfo{publisher}{Springer-Verlag}, \bibinfo{year}{1980}).

\bibitem[{\citenamefont{Walet et~al.}(1991)\citenamefont{Walet, DoDang, and
  Klein}}]{Walet}
\bibinfo{author}{\bibfnamefont{N.~R.} \bibnamefont{Walet}},
  \bibinfo{author}{\bibfnamefont{G.}~\bibnamefont{DoDang}}, \bibnamefont{and}
  \bibinfo{author}{\bibfnamefont{A.}~\bibnamefont{Klein}},
  \bibinfo{journal}{Phys.\ Rev.} \textbf{\bibinfo{volume}{C43}},
  \bibinfo{pages}{2254} (\bibinfo{year}{1991}).

\bibitem[{\citenamefont{Kohmura et~al.}(2000)\citenamefont{Kohmura, Hashimoto,
  Ohta, and Maruyama}}]{Kohmura2}
\bibinfo{author}{\bibfnamefont{T.}~\bibnamefont{Kohmura}},
  \bibinfo{author}{\bibfnamefont{Y.}~\bibnamefont{Hashimoto}},
  \bibinfo{author}{\bibfnamefont{H.}~\bibnamefont{Ohta}}, \bibnamefont{and}
  \bibinfo{author}{\bibfnamefont{M.}~\bibnamefont{Maruyama}},
  \bibinfo{journal}{Phys.\ Rev.} \textbf{\bibinfo{volume}{C61}},
  \bibinfo{pages}{034315} (\bibinfo{year}{2000}).

\bibitem[{\citenamefont{Lederer and Shirley}(1978)}]{Lederer}
\bibinfo{editor}{\bibfnamefont{C.~M.} \bibnamefont{Lederer}} \bibnamefont{and}
  \bibinfo{editor}{\bibfnamefont{V.~S.} \bibnamefont{Shirley}}, eds.,
  \emph{\bibinfo{title}{Tables of Isotopes}} (\bibinfo{publisher}{Wiley
  Interscience}, \bibinfo{year}{1978}).

\bibitem[{\citenamefont{Ohta}(2000)}]{Ohta}
\bibinfo{author}{\bibfnamefont{H.}~\bibnamefont{Ohta}}, Master's thesis,
  \bibinfo{school}{Univ. of Tsukuba} (\bibinfo{year}{2000}).

\bibitem[{\citenamefont{Barranco et~al.}(1988)\citenamefont{Barranco, Broglia,
  and Bertsch}}]{Bertsch}
\bibinfo{author}{\bibfnamefont{F.}~\bibnamefont{Barranco}},
  \bibinfo{author}{\bibfnamefont{R.~A.} \bibnamefont{Broglia}},
  \bibnamefont{and} \bibinfo{author}{\bibfnamefont{G.~F.}
  \bibnamefont{Bertsch}}, \bibinfo{journal}{Phys. Rev. Lett.}
  \textbf{\bibinfo{volume}{60}}, \bibinfo{pages}{507} (\bibinfo{year}{1988}).

\bibitem[{\citenamefont{Barranco et~al.}(1990)\citenamefont{Barranco, Bertsch,
  Broglia, and Vigezzi}}]{Bertsch2}
\bibinfo{author}{\bibfnamefont{F.}~\bibnamefont{Barranco}},
  \bibinfo{author}{\bibfnamefont{G.~F.} \bibnamefont{Bertsch}},
  \bibinfo{author}{\bibfnamefont{R.~A.} \bibnamefont{Broglia}},
  \bibnamefont{and} \bibinfo{author}{\bibfnamefont{E.}~\bibnamefont{Vigezzi}},
  \textbf{\bibinfo{volume}{A512}}, \bibinfo{pages}{253} (\bibinfo{year}{1990}).

\bibitem[{\citenamefont{Elliott}(1958)}]{Elliott}
\bibinfo{author}{\bibfnamefont{J.~P.} \bibnamefont{Elliott}},
  \bibinfo{journal}{Proc. Royal. Soc.} \textbf{\bibinfo{volume}{A245}},
  \bibinfo{pages}{128} (\bibinfo{year}{1958}).

\bibitem[{\citenamefont{Lichtenberg}(1970)}]{Lichtenberg}
\bibinfo{author}{\bibfnamefont{D.~B.} \bibnamefont{Lichtenberg}},
  \emph{\bibinfo{title}{Unitary Symmetry and Elementary Particles}}
  (\bibinfo{publisher}{Academic Press}, \bibinfo{year}{1970}).

\bibitem[{\citenamefont{Arve et~al.}(1987)\citenamefont{Arve, Bertsch, Negele,
  and Puddu}}]{Arve}
\bibinfo{author}{\bibfnamefont{P.}~\bibnamefont{Arve}},
  \bibinfo{author}{\bibfnamefont{G.~F.} \bibnamefont{Bertsch}},
  \bibinfo{author}{\bibfnamefont{J.~W.} \bibnamefont{Negele}},
  \bibnamefont{and} \bibinfo{author}{\bibfnamefont{G.}~\bibnamefont{Puddu}},
  \bibinfo{journal}{Phys.\ Rev.} \textbf{\bibinfo{volume}{C36}},
  \bibinfo{pages}{2018} (\bibinfo{year}{1987}).

\bibitem[{\citenamefont{Kohmura et~al.}(2002)\citenamefont{Kohmura, Maruyama,
  Ohta, and Hashimoto}}]{kohmura3}
\bibinfo{author}{\bibfnamefont{T.}~\bibnamefont{Kohmura}},
  \bibinfo{author}{\bibfnamefont{M.}~\bibnamefont{Maruyama}},
  \bibinfo{author}{\bibfnamefont{H.}~\bibnamefont{Ohta}}, \bibnamefont{and}
  \bibinfo{author}{\bibfnamefont{Y.}~\bibnamefont{Hashimoto}},
  \bibinfo{journal}{Prog.\ Theor.\ Phys.} \textbf{\bibinfo{volume}{107}},
  \bibinfo{pages}{87} (\bibinfo{year}{2002}).

\end{thebibliography}

%
%
%
%
%
%
%
%
%
%
%
%
%

\end{document}